    \documentclass{jpp}

    \shorttitle{Electron runaway in ASDEX Upgrade experiments of varying core temperature}
    \shortauthor{O. Linder, G. Papp, E. Fable, F. Jenko, G. Pautasso and others}

    \title{Electron runaway in ASDEX Upgrade experiments of varying core temperature}
    \author{%
        O. Linder\aff{1}
            \corresp{\email{\href{mailto:Oliver.Linder@ipp.mpg.de}{Oliver.Linder@ipp.mpg.de}}},
        G. Papp\aff{1},
        E. Fable\aff{1},
        F. Jenko\aff{1},
        G. Pautasso\aff{1},
        the ASDEX Upgrade Team
            \corresp{See author list of H. Meyer \textit{et al} 2019 %
                \textit{Nucl.~Fusion} \textbf{59}, %
                \href{https://doi.org/10.1088/1741-4326/ab18b8}{112014}}
        \and
        the EUROfusion MST1 Team%
            \corresp{See author list of B. Labit \textit{et al} 2019 %
                \textit{Nucl.~Fusion} \textbf{59}, %
                \href{https://doi.org/10.1088/1741-4326/ab2211}{086020}}
        }

    \affiliation{%
        \aff{1}Max-Planck-Institut f\"ur Plasmaphysik, 85748 Garching, Germany
        }

    \usepackage{graphicx}

    \usepackage[utf8]{inputenc}
    \usepackage[T1]{fontenc}

    \usepackage{xcolor}

    \usepackage[
        pdftex,
        pdffitwindow    =true,
        pdftitle        ={Electron runaway in ASDEX Upgrade experiments of varying core temperature},
        pdfsubject      ={High-temperature plasma physics},
        pdfauthor       ={O. Linder, G. Papp, E. Fable, F. Jenko, G. Pautasso, 
                            the ASDEX Upgrade Team, and the EUROfusion MST1 Team},
        pdfcreator      ={O. Linder, G. Papp, E. Fable, F. Jenko, G. Papp, G. Pautasso, 
                            the ASDEX Upgrade Team, and the EUROfusion MST1 Team},
        pdfproducer     ={O. Linder, G. Papp, E. Fable, F. Jenko, G. Papp, G. Pautasso, 
                            the ASDEX Upgrade Team, and the EUROfusion MST1 Team},
        colorlinks      =true,
        linkcolor       ={hyperlinkcolor!75!black},
        citecolor       ={hyperlinkcolor!75!black},
        urlcolor        ={hyperlinkcolor!75!black},
        pdfkeywords     ={tokamaks, disruptions, massive gas injection, 
        					runaway electrons},
        ]{hyperref}

    \usepackage{amsmath}
    \usepackage{upgreek}

    \usepackage{blindtext}
    
    \usepackage{tikz}
    	\usetikzlibrary{backgrounds,calc,external,math,decorations.text}
    \usepackage{pgfplots}
    	\usepgfplotslibrary{fillbetween,groupplots}
    
    \definecolor{hyperlinkcolor}{rgb}{0.,0.,1.}
	
	\definecolor{set1_1}{HTML}{e41a1c}	
	\definecolor{set1_2}{HTML}{377eb8}	
	\definecolor{set1_3}{HTML}{4daf4a}	

\begin{document}
    
    \maketitle

\begin{abstract}
The formation of a substantial post-disruption runaway electron current in ASDEX Upgrade material injection experiments is determined by avalanche multiplication of a small seed population of runaway electrons. For the investigation of these scenarios, the runaway electron description of the coupled 1.5D transport solvers \texttt{ASTRA-STRAHL} is amended by a fluid-model describing electron runaway caused by the hot-tail mechanism. Applied in simulations of combined background plasma evolution, material injection, and runaway electron generation in ASDEX Upgrade discharge \#33108, both the Dreicer and hot-tail mechanism for electron runaway produce only $\sim 3~{\rm kA}$ of runaway current. In colder plasmas with core electron temperatures $T_{\rm e,c}$ below 9~keV, the post-disruption runaway current is predicted to be insensitive to the initial temperature, in agreement with experimental observations. Yet in hotter plasmas with $T_{\rm e,c} > 10$~keV, hot-tail runaway can be increased by up to an order of magnitude, contributing considerably to the total post-disruption runaway current. In ASDEX Upgrade high temperature runaway experiments, however, no runaway current is observed at the end of the disruption, despite favourable conditions for both primary and secondary runaway.
\end{abstract}


\section{Introduction}
\label{sec_1}
In future current-carrying fusion devices, the formation of a substantial population of runaway electrons during the sudden loss of thermal confinement poses a significant threat to the integrity of the plasma vessel. Already in present-day devices, beams of lost runaway electrons are observed to damage plasma facing components, e.g. at JET \citep{Matthews16} or at Alcator C-Mod \citep{Tinguely18}. However in high-current devices, a larger runaway current is expected due to increased avalanche multiplication \citep{Boozer19}. As the total energy carried by a runaway beam grows quadratically with the runaway current \citep{Martin-Solis14}, the threat to high current fusion devices is amplified. Therefore, runaway electron generation has to be suppressed and potential disruptions mitigated \citep{Breizman19}. 

Suppression of electron runaway may be achieved through massive material injection, as proposed for ITER \citep{Lehnen15}. This scheme is currently being investigated across several machines using massive gas injection (MGI), e.g. at ASDEX Upgrade (AUG) \citep{Pautasso17,Pautasso20} or TCV \citep{Coda19}, and shattered pellet injection (SPI), e.g. at DIII-D \citep{Commaux10,PazSoldan20} and JET \citep{Reux21}. Given the unfavourable scaling of the runaway electron threat to future devices due to increased avalanche multiplication, experimental investigation is complemented by theoretical and computational studies to aid in extrapolation from present to future devices \citep{Breizman19}.

Owing to the complexity of the runaway electron problem, different computational tools are used for the investigation of different aspects of electron runaway. The most accurate description is achieved by kinetic tools, such as e.g. the full-$f$ Fokker-Planck solver \texttt{CODE} \citep{Stahl16}, where the runaway fluxes are determined through evolution of the momentum-space electron distribution. However, the simultaneous spatio-temporal evolution of the background plasma or impurities injected is challenging to calculate in these frameworks \citep{Hoppe21}. For this purpose, 1D transport codes such as e.g. \texttt{ASTRA-STRAHL} \citep{Fable13,Dux99,Linder20} or \texttt{GO} \citep{Papp13,Vallhagen20} can be applied. Here, electron runaway is described through a fluid treatment, as a kinetic description greatly increases the computational cost. For a description of the 3D spatio-temporal evolution of the magnetic field during disruptions, non-linear MHD codes such as e.g. \texttt{JOREK} \citep{Bandaru19} are used.

The recent development of sophisticated reduced kinetic models describing electron runaway due to momentum-space diffusion of thermal electrons \citep{Hesslow19_2} and knock-on collisions of existing runaways with the thermal bulk \citep{Hesslow19} has accelerated modelling efforts. Applied inside the transport code \texttt{ASTRA-STRAHL}, simulations of the spatio-temporal evolution of runaway electron population, background plasma, and material injected have recently been found capable of describing AUG disruptions, as demonstrated modelling AUG discharge \#33108 \citep{Linder20}. 

In this work, we investigate runaway electron (seed) generation in AUG experiments of varying core temperature between 4~keV and 20~keV by means of \texttt{ASTRA-STRAHL} simulations. For this purpose, we expand upon the findings by \citet{Linder20}, performing coupled simulations of background plasma evolution, material injection and electron runaway. As kinetic modelling using \texttt{CODE} suggests formation of a seed population of runaway electrons predominantly due to rapid cooling \citep{Insulander20,Hoppe21}, the runaway electron generation models used in \texttt{ASTRA-STRAHL} are amended by a model by \citet{Smith08} describing this effect. The tool-kit \texttt{ASTRA-STRAHL} is then applied for the investigation of the (seed) runaway electron population in simulations of AUG discharge \#33108. Throughout the simulations performed, the pre-injection on-axis electron temperature is varied between 4~keV and 20~keV, as SPI experiments in DIII-D suggest a growing seed runaway population as the electron temperature increases \citep{PazSoldan20}. The simulation results obtained are compared against measurements of AUG disruption experiments.

This paper is organized as follows. A brief description of the model employed is provided in section~\ref{sec_2}, with experimental aspects of AUG discharge \#33108 covered in section~\ref{sec_3}. More details on both parts can be found in \citet{Linder20}. Simulations of runaway electron generation in the AUG discharge chosen are presented in section~\ref{sec_4}. The impact of a variation of the pre-injection on-axis electron temperature on the post-disruption runaway electron current calculated is discussed in section~\ref{sec_5}. Finally, a conclusion is provided in section~\ref{sec_6}. Additionally, a simplified model for the hot-tail runaway electron current density at the end of the thermal quench is presented in appendix~\ref{sec_A}; the impact of the average runaway electron velocity on the post-disruption runaway current discussed in appendix~\ref{sec_B}.

\section{Model description}
\label{sec_2}
The spatio-temporal evolution of the main tokamak plasma, material introduced through massive gas injection (MGI), and runaway electrons generated in the process can be described by the coupled 1.5D transport codes \texttt{ASTRA} \citep{Fable13} and \texttt{STRAHL} \citep{Dux99}. The suitability of this toolkit for the simulation of runaway electron generation during MGI has recently been demonstrated by \citet{Linder20}. Building on the model presented, the capabilities of \texttt{ASTRA-STRAHL} are expanded to additionally consider electron runaway due to the hot-tail mechanism. Therefore, only a brief overview of \texttt{ASTRA-STRAHL} is given, with details described in \citet{Linder20}.

\subsection{The coupled transport codes \texttt{ASTRA-STRAHL}}
\label{sec_2.1}
The evolution of the main plasma and impurity species introduced is calculated by \texttt{ASTRA} and \texttt{STRAHL}, respectively, following the macroscopic transport equation
\begin{align}
    \label{eq_2.1}
	\frac{\partial Y}{\partial t} = \left(\frac{\partial V}{\partial \rho}\right)^{-1} \frac{\partial}{\partial \rho} \left( \frac{\partial V}{\partial \rho} \left\langle \left( \Delta \rho \right)^2 \right\rangle \left\{ D \frac{\partial Y}{\partial \rho} - v~Y \right\} \right) + \sum_j S_j 
\end{align}
for a fluid quantity $Y$ in the presence of diffusion $D$, convection $v$, and sources $S_j$. The quantity $\rho$ denotes the toroidal flux-surface label, with $V$ being the flux-surface volume. 

Inside \texttt{ASTRA}, the poloidal magnetic flux $\Uppsi$, both the electron temperature $T_{\rm e}$ and ion temperature $T_{\rm i}$, and the density $n_{\rm RE}$ of runaway electrons are evolved. In the case of electron heat transport, sources $S_j$ due to Ohmic heating, electron-to-ion heat transport and impurity radiation from \texttt{STRAHL} (line radiation and Bremsstrahlung) are taken into account throughout the entirety of the simulations and assumed to outweigh radial transport \citep{Feher11} (confirmed by the simulations presented). Consequently, turbulent radial transport is neglected. The electron density $n_{\rm e}$ is calculated from quasi-neutrality, i.e. $n_{\rm e}(t) = n_{\rm D} + \sum_k \left\langle Z_k \right\rangle n_k(t)$ with densities $n_k$ and average charges $\left\langle Z_k \right\rangle$ of the impurities $k$ evolved by \texttt{STRAHL}. The magnetic equilibrium is obtained from the \texttt{ASTRA} built-in 3-moment solver, applicable for circular discharges of MGI experiments in AUG \citep{Pautasso17,Pautasso20}.

The impurity densities $n_{k,i}$ are evolved by \texttt{STRAHL} for each charge state $i$ under consideration of electron impact ionization and recombination rates from \texttt{ADAS} \citep{Summers04}. Neutrals originating from a gas valve are deposited in the simulation domain just outside the last closed flux surface (LCFS) and propagate into the core plasma with thermal velocity $v_{k,0} = v_{\rm th} = \sqrt{T/m}$. The source strength $-{\rm d}N_k/{\rm d}t$ is determined from the continuity equation ${\rm d}N_k/{\rm d}t + v_{k,0} N_k A_{\rm v}(t)/V_{\rm v} = 0$ for a valve with particle inventory $N_k$, aperture size $A_{\rm v}(t)$, and volume $V_{\rm v}$. Impurity transport due to neoclassical processes is described by \texttt{NEOART} \citep{Peeters00}. 

Following the injection of impurities, $(2,1)$ magnetohydrodynamic (MHD) modes and higher harmonics are triggered as the cold gas front reaches the $q=2$ surface \citep{Fable16} at time $t_{q=2}$. As a result, the current density inside the $q=2$ surface is redistributed, which is achieved in the simulations by flattening the $q$-profile to $q=2$ under conservation of the total poloidal magnetic flux. During the break-up of the magnetic surfaces, the transport of ionized material and heat is greatly enhanced until closed flux surfaces have re-emerged. To mimic this effect inside \texttt{ASTRA}, additional transport coefficients of the form
\begin{align}
    \label{eq_2.2}
	X_{\rm add}(t) = X_{\rm add}^{\rm max}~\exp\left( - \frac{t - t_{q=2}}{\tau_{\rm add}} \right) \Uptheta(t - t_{q=2})
\end{align}
are applied for both diffusive and convective transport inside the $q=2$ surface with $D_{\rm add}^{\rm max} = 100~{\rm m}^2/{\rm s}$, $v_{\rm add}^{\rm max} = -1000~{\rm m}/{\rm s}$, $\chi_{\rm add}^{\rm max} = 100~{\rm m}^2/{\rm s}$ \citep{Feher11} and $\tau_{\rm add} = 1.0~{\rm ms}$. The evolution of plasma parameters in \texttt{ASTRA-STRAHL} simulations applying this approach for discharge AUG \#33108 studied in this work has been compared in detail to experimental observations in the publication by \citet{Linder20}, where application of these coefficients was found necessary to reproduce the experimentally observed increase of the line averaged electron density. Please note, that in this work, the additional transport coefficients are set to generic values (instead of a refined fit) as experimental observations are adequately described under a moderate variation of these coefficients by up to 50\%. In the simulations of varying pre-injection on-axis electron temperature between 4~keV and 20~keV discussed in section~\ref{sec_5}, transport coefficients of identical magnitude are prescribed, since the MHD modes triggered are largely current driven. As such, a (strong) dependence of the mode amplitude on pre-disruption temperature and pressure is not expected. A more detailed investigation on this subject is planned for future work.

The simulations presented in this work are carried out employing a radial grid of 401 points inside \texttt{ASTRA}, extending from the magnetic axis up to the LCFS. For \texttt{STRAHL} calculations, the grid is expanded to additionally include the scrape-off layer. Both minimum and constant time step in \texttt{ASTRA} and \texttt{STRAHL}, respectively, are set to $10^{-5}~$ms to resolve transient events. The suitability of these simulation settings was ensured by means of convergence scans of radial and temporal resolution in pre-study simulations.

\subsection{Runaway electron generation}
\label{sec_2.2}
The process of electron runaway is described by reduced fluid-models, providing sources $S_j$ for the evolution of the runaway electron density $n_{\rm RE}$ inside \texttt{ASTRA}%
\footnote{The runaway electron generation models discussed are implemented as a standalone \texttt{Fortran} module, available at \href{https://github.com/o-linder/runawayelectrongeneration}{https://github.com/o-linder/runawayelectrongeneration}.}%
. Mechanisms for runaway electron generation considered in this work include hot-tail generation due to rapid cooling (see section~\ref{sec_2.2.1}), Dreicer generation due to momentum-space diffusion of thermal electrons (see section~\ref{sec_2.2.2}), and avalanche generation due to knock-on collision of existing runaway with thermal electrons (see section~\ref{sec_2.2.3}). Further mechanisms due to nuclear processes \citep{Vallhagen20} are not taken into account given AUG's non-nuclear environment. Feedback of the runaway electron population on the poloidal magnetic flux evolution $\Uppsi(t)$ is considered by adding the runaway electron current density to the total plasma current density under the assumption $\left\langle v_{\rm RE} \right\rangle = c$.

\subsubsection{Hot-tail generation}
\label{sec_2.2.1}
In events of rapid plasma cooling, as in the case of tokamak disruptions, electron runaway may occur. Under these conditions, the high-energy tail of the electron energy distribution function equilibrates slower than the thermal bulk and may thus exceed the critical energy for runaway \citep{Chiu98,Harvey00}. The runaway electron population generated due to this process can be described by reduced fluid models, e.g. by the work of \citet{Smith08} and \citet{Feher11}. However, compared to kinetic simulations with the full-$f$ continuum Fokker-Planck solver \texttt{CODE} \citep{Stahl16}, these reduced models are found to underestimate the hot-tail density by up to an order of magnitude as the impact of the electric field on the underlying electron distribution function is not taken into account by these models \citep{Harvey19,Breizman19}. Simultaneously, the computational cost of kinetic solvers renders application in transport simulations impractical. Therefore, cheaper and more accurate models are currently being developed by \citet{Svenningsson20}, which however are not available yet for practical applications with varying effective plasma charge $Z_{\rm eff}$. For this reason, the model by \citet{Smith08} is applied in this work for the calculation of the hot-tail runaway electron population. Note, that in a recent validation of this model by \citet{Petrov21} with the Fokker-Planck solver \texttt{CQL3D}, an additional $Z_{\rm eff}$ dependent factor of order unity was proposed for the definition of the critical velocity.

According to the model by \citet{Smith08}, the hot-tail runaway electron density $n_{\rm hot}$ at time $t$ is obtained from the velocity-space integral across the runaway region as
\begin{align}
    \label{eq_2.3}
    n_{\rm hot}(t) &= \frac{4 n_{{\rm e},0}}{\sqrt{\upi} v_{{\rm th},0}^3} \int_{v_{\rm c}(t)}^{\infty} \left( v^2 - v_{\rm c}(t)^2 \right) \exp{\left( - \left[ \frac{v^3}{v_{{\rm th},0}^3} + 3 \tau(t) \right]^{2/3} \right)} {\rm d}v ~,
\end{align}
where $v_{\rm th}$ denotes the thermal velocity, $v_{\rm th}^2 = 2 T_{\rm e}/m_{\rm e}$, and $v_{\rm c}$ the critical velocity for electron runaway, $v_{\rm c}^2 = e^3 n_{\rm e} \ln \Lambda/4 \upi \varepsilon_0^2 m_{\rm e} E_\parallel$ with $\ln \Lambda$ being the Coulomb logarithm for thermal-thermal collisions, i.e. $\ln \Lambda = 16.1 - 0.5 \log{\left(n_{\rm e}/10^{19}~{\rm m}^{-3}\right)} + \log{\left( T_{\rm e}/{\rm keV} \right)}$. Quantities evaluated at the onset of rapid cooling are denoted by indices "$0$". The parameter $\tau(t)$ is a normalized time, i.e. $\tau(t) = \nu_0 \int_{t_0}^t n_{\rm e}(\tilde{t})~{\rm d}\tilde{t}/n_{{\rm e},0}$, with the thermal-thermal collision frequency $\nu = n_{\rm e}e^4\ln\Lambda/4\upi \varepsilon_0^2 m_{\rm e}^{2} v_{\rm th}^{3}$. 

The expression~\eqref{eq_2.3} introduced by \citet{Smith08} for the hot-tail density assumes an instantaneous drop of the electron temperature from $T_{\mathrm{e},0}$ to the final temperature $T_{\rm e,fin}$. However motivated by an exponential decay of the temperature due to plasma cooling \citep{Smith08}, the hot-tail density evolution under assumption of an exponential decay of the temperature, i.e.
\begin{align}
    \label{eq_2.4}
	T_{\rm e}(t) = (T_{\rm e,0} - T_{\rm e,fin}) \exp\left( - \frac{t-t_0}{t_{\rm dec}} \right) + T_{\rm e,fin} ~,
\end{align}
can be described by modifying the expression for $\tau(t)$. In the work by \citet{Smith08}, the temporal evolution of this parameter is obtained through numerical integration of a high moment of the kinetic equation for a two-component distribution function. In the case of an exponential electron density evolution, the numerical solution obtained for $\tau(t)$ is well approximated by $\tau(t) = \nu_0(t - t_0 - t_{\rm dec}) \Uptheta(t - t_0 - t_{\rm dec}) n_{\rm e,fin}/n_{{\rm e},0}$ for $t - t_0 > 3~t_{\rm dec}$ \citep{Smith08}. However to describe $\tau(t)$ more accurately during the initial phase of rapid cooling, an alternative expression is introduced and used throughout this work
\begin{align}
    \label{eq_2.5}
    \tau(t) &= \nu_0 \frac{n_{\rm e,fin}}{n_{{\rm e},0}}
        \begin{cases}
            \frac{(t-t_0)^2}{4 t_{\rm dec}}~, &\qquad t - t_0 < 2 t_{\rm dec} \\
                \left( t - t_0 - t_{\rm dec} \right) ~, &\qquad t - t_0 \geq 2 t_{\rm dec}
        \end{cases} ~.
\end{align}
For evaluation of the hot-tail density, a closed form of expression \eqref{eq_2.3} cannot be provided, necessitating numerical integration. Inside \texttt{ASTRA}, the integral is evaluated using Kepler's rule as the integrand falls of monotonically and sufficiently fast for $v \to \infty$. The hot-tail runaway electron density $n_{\rm hot}$ obtained can be used inside \texttt{ASTRA} directly for subsequent calculations of the runaway electron current density and secondary runaway generation, eliminating the necessity to evaluate the macroscopic transport equation \eqref{eq_2.1} for the hot-tail population. However importantly, evaluation of the instantaneous hot-tail population through equation \eqref{eq_2.3} requires characterisation of parameters at onset and end of the thermal quench, being the time $t_0$ of the onset of the thermal quench, the electron temperature $T_{\rm e}(t_0)$, the temperature decay time scale $t_{\rm dec}$, as well as the electron density at onset and end of the thermal quench, i.e. $n_{\rm e}(t_0)$ and $n_{\rm e}(t_{\rm fin})$. As onset and end of the thermal quench cannot be determined during a simulation, the required parameters are calculated in post-simulation analysis (see section~\ref{sec_2.3}) and applied in a subsequent simulation for the calculation of the hot-tail population. Convergence of the parameters obtained has to be assessed and simulations repeated if convergence is not met. Note, that in the simulations presented in this work, one iteration to determine the thermal quench parameters was sufficient to achieve an averaged iteration accuracy of less than 1\% across all grid points, being in all cases less than 5\%.

\subsubsection{Dreicer generation}
\label{sec_2.2.2}
The process of electron runaway due to momentum-space diffusion of thermal electrons in the presence of partially ionized mid- to high-$Z$ impurities cannot be described by analytical reduced fluid models, as a result of the complicated energy dependence of collision frequencies at near-thermal energies \citep{Hesslow19_2}. Analytical expressions under consideration of fully ionized impurities only \citep{Connor75} have been demonstrated to overestimate electron runaway under certain conditions \citep{Hesslow19_2}. Applied in transport simulations of runaway electron generation during MGI, a noticeably increased seed population in contrast to experimental observations is obtained \citep{Linder20}. Therefore, instead of reduced fluid models, a neural network model for the calculation of Dreicer growth rates \citep{Hesslow19_2} is utilized in this work.

The neural network by \citet{Hesslow19_2} is based on simulations of \texttt{CODE}. Training of the network was performed with argon and neon impurities, generalized for application to other species using 8 input parameters $\boldsymbol{x}$. The Dreicer source rate $S_{\rm D}$ is thus obtained through evaluation of
\begin{align}
    \label{eq_2.6}
	S_{\rm D} &= \nu n_{\rm e} \exp \left( \mathcal{F}(\boldsymbol{W}_5, \mathcal{F}(\mathsfbi{W}_4, \mathcal{F}(\mathsfbi{W}_3, \mathcal{F}(\mathsfbi{W}_2, \mathcal{F}(\mathsfbi{W}_1, \boldsymbol{x}, \boldsymbol{b}_1), \boldsymbol{b}_2), \boldsymbol{b}_3), \boldsymbol{b}_4), b_5) \right) ~, \\
	\label{eq_2.7}
	\mathcal{F}(W, x, b) &= \tanh \left( W x + b \right) ~,
\end{align}
with weights $\mathsfbi{W}_i$ and biases $\boldsymbol{b}_i$ (see \citet{Hesslow19_2} for details).

\subsubsection{Avalanche generation}
\label{sec_2.2.3}
The generation of secondary runaway electrons due to knock-on collisions of existing runaways with thermal electrons in the presence of partially ionized impurities can be described through a reduced fluid model by \citet{Hesslow19}. The avalanche source rate $S_{\rm av}$ is calculated from			
\begin{align}
    \label{eq_2.8}
	S_{\rm av} &= n_{\rm RE} \frac{e}{m_{\rm e}c \ln \Lambda_{\rm c}} \frac{n_{\rm e}^{\rm tot}}{n_{\rm e}} \frac{E_{\parallel} - E_{\rm c}^{\rm eff}}{\sqrt{4 + \bar{\nu}_{\rm slow}(p_\star) \bar{\nu}_{\rm defl}(p_\star)}} ~,
\end{align}
with the relativistic Coulomb logarithm $\ln \Lambda_{\rm c} = \ln \Lambda - 0.5 \ln (T/m_{\rm e}c^2)$. The total electron density $n_{\rm e}^{\rm tot}$ comprises both free plasma electrons $n_{\rm e}$ and electrons bound to impurity ions. In the presence of partially ionized impurities, the critical electric field $E_{\rm c} = n_{\rm e}e^3\ln \Lambda_c/4\upi \varepsilon_0^2 m_{\rm e}c^2$ for runaway is increased, the effect of which is described by the effective critical electric field $E_{\rm c}^{\rm eff}$ defined in \citet{Hesslow18}. Expressions for the slowing-down frequency $\bar{\nu}_{\rm slow}$ and for the generalized deflection frequency $\bar{\nu}_{\rm defl}$ are found in \citet{Hesslow18_2} and \citet{Hesslow18}. Noticeably, the effective critical momentum $p_\star$ depends on both frequencies through $p_\star = \sqrt[4]{\bar{\nu}_{\rm slow}(p_\star) \bar{\nu}_{\rm defl}(p_\star)}/\sqrt{E_\parallel/E_{\rm c}}$, thus requiring numerical evaluation of these parameters.

\begin{figure}
    \centering
    \includegraphics[width=132.6444mm]{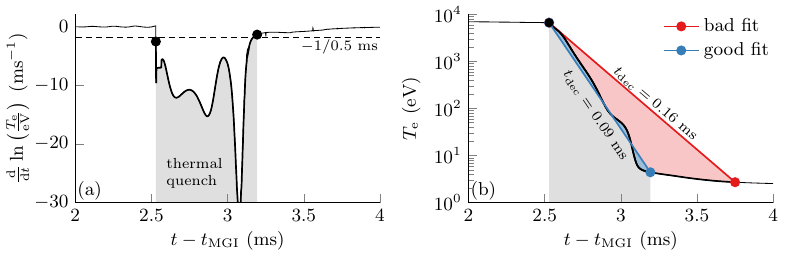}
    \caption{\label{fig_1}%
        The occurrence of a thermal quench is determined from the electron temperature evolution. (a) The temporal derivative of the logarithmic temperature falling below a threshold of $-1/0.5~{\rm ms}$ marks the onset of the quench. (b) The end is obtained from an exponential fit of the electron temperature, illustrated for both a suitable (blue) and a poor (red) choice of the decay time scale $t_{\rm dec}$.
        }
\end{figure}

\subsection{Determining thermal quench parameters}
\label{sec_2.3}
The calculation of characteristic quantities of the thermal quench for the evaluation of the hot-tail runaway population during the thermal quench (see section~\ref{sec_2.2.1}) is performed in post-simulation analysis. Onset $t_0$ and end $t_{\rm fin}$ of the thermal quench are determined from the electron temperature evolution. The required values for both the electron temperature and density are then obtained through evaluation of these quantities at $t_0$ and $t_{\rm fin}$, respectively.

The time $t_0$ of the onset of the thermal quench is defined as the time when the instantaneous logarithmic temperature change ${\rm d}\ln(T_{\rm e}(t)/{\rm eV})/{\rm d}t$ falls below a threshold value $-1/\tilde{t}_{\rm dec} = -1/0.5~{\rm ms}$, with $\tilde{t}_{\rm dec}$ being the instantaneous temperature decay time (see figure~\ref{fig_1}(a)). Both the end $t_{\rm fin}$ of the thermal quench and the temperature decay time scale $t_{\rm dec}$ are determined through a linear fit $\ln \tilde{T}_{\rm e}(t)$ of the logarithmic electron temperature evolution (see figure~\ref{fig_1}(b)). Under the assumption $T_{\rm e}(t_0) \gg T_{\rm e}(t_{\rm fin})$, the ansatz for $T_{\rm e}(t)$ of equation~\eqref{eq_2.4} can be reduced to $\ln \left( T_{\rm e}(t)/T_{\rm e}(t_0)\right) = -\left( t - t_0 \right)/t_{\rm dec}$, thus yielding the time scale $t_{\rm dec}$. The end of the thermal quench is defined as the last time point where $\tilde{T}_{\rm e}(t_{\rm fin}) = T_{\rm e}(t_{\rm fin})$, i.e. before the fit falls off below the actual temperature. The quality of the fit is evaluated for $t \in [t_0, t_{\rm fin}]$.
    
\section{ASDEX Upgrade runaway electron experiments}
\label{sec_3}
\subsection{Reference scenario}
\label{sec_3.1}
Simulations of runaway electron generation are performed for artificially disrupted ASDEX Upgrade experiments through MGI \citep{Pautasso17,Pautasso20}. The plasma parameters chosen in this work are based on ASDEX Upgrade discharge \#33108 (for details see \citet{Linder20}). In this experiment, argon (Ar) was injected at $t_{\rm inj} = 1.0~{\rm s}$ after breakdown from a gas valve of volume $100~{\rm cm}^3$ and initial Ar pressure of $0.73~{\rm bar}$ into an L-mode limiter plasma with low average electron density of $\left\langle n_{\rm e} \right\rangle = 2.8 \times 10^{19}~{\rm m}^{-3}$ and high peaked electron temperature of $T_{\rm e}(\rho = 0) = 9.3~{\rm keV}$ at the magnetic axis. A peaked temperature profile is achieved through application of 2.6~MW of on-axis electron cyclotron resonance heating (ECRH) during the last 0.1~s prior to MGI. As a result of Ar injection, the plasma stored energy is removed through impurity radiation and the plasma current decreases from initially $763~{\rm kA}$ down to $225~{\rm kA}$, carried by relativistic electrons. Additional characteristic parameters of AUG \#33108, as well as criteria for selecting similar runaway electron shots for analysis in section~\ref{sec_5.5}, are listed in table~\ref{table_1}.

\begin{table}
  \begin{center}
  \begin{tabular}{lcc}
			Quantity & AUG \#33108 & similar shots \\ [3pt]
		$I_{\rm p,0}~({\rm MA})$ & 0.76 & $0.60 - 0.90$ \\
		$p_{\rm Ar}~({\rm bar})$ & 0.73 & $0.60 - 0.85$ \\
		$B_{\rm tor}~({\rm T})$ & 2.50 & $2.30 - 2.70$ \\
		$q_{95}$ & $3.79$ & $3.50 - 4.10$
  \end{tabular}
  \caption{Characteristic parameters for runaway electron experiments in AUG, being the pre-disruptive plasma current $I_{\rm p,0}$, the valve Ar pressure $p_{\rm Ar}$, the toroidal magnetic field $B_{\rm tor}$, and the edge safety factor $q_{95}$. Values for the reference discharge AUG \#33108 are given, as well as criteria for selecting similar shots from all AUG runaway electron experiments performed.}
  \label{table_1}
  \end{center}
\end{table}

\subsection{Gaussian process regression for experimental fitting}
\label{sec_3.2}
Reconstruction of experimental profiles often requires fitting of measured data. Application of a probabilistic approach under consideration of uncertainties allows a reliable estimate of experimental quantities. Therefore, Gaussian process regression (GPR) techniques are employed in this work through application of a tool-set by \citet{Ho19}, based on work by \citet{Chilenski15}. Using these tools, reconstruction of pre-injection electron temperature profiles $T_{\rm e}(t_{\rm inj},\rho)$ from electron cyclotron emission (ECE) and Thomson scattering (TS) measurements is performed in sections~\ref{sec_5.1} for the entire plasma radius and in section~\ref{sec_5.5} for a better reconstruction of the on-axis value $T_{\rm e}(t_{\rm inj},0)$. An estimation of the experimentally measured runaway electron current as a function of $T_{\rm e}(t_{\rm inj},0)$ is also performed using GPR.

Applying Bayesian probability theory, robust reconstruction of these profiles, as well as of associated gradients and uncertainties, is performed from covariance functions $k(x,x')$ utilizing normally distributed weights. Where stated in this work, profile estimation through GPR is performed using a rational quadratic covariance function
\begin{align}
    \label{eq_3.1}
	k(x,x') &= \sigma^2 \left( 1 + \frac{(x-x')^2}{2\alpha l^2} \right)^\alpha 
\end{align}
with characteristic length scale $l$. When simpler estimates are sufficient, plasma profiles are instead reconstructed using an mtanh function \citep{Schneider12}.

\section{Electron runaway in ASDEX Upgrade \#33108}
\label{sec_4}
Coupled transport simulations of Ar injection, background plasma evolution, and RE generation are performed with \texttt{ASTRA-STRAHL} for AUG discharge \#33108. The evolution of the Ar-induced disruption throughout the simulation is described in section~\ref{sec_4.1}. The generation of a seed population is discussed in section~\ref{sec_4.2}, whereas the avalanche multiplication following is covered in section~\ref{sec_4.3}. The spatio-temporal evolution of the runaway electron current density contributions and the Ohmic current density is illustrated, in addition to this paper, in a supplementary movie available at \href{https://arxiv.org/abs/2101.04471}{https://arxiv.org/abs/2101.04471}.

\subsection{Simulation of thermal and current quench}
\label{sec_4.1}
Simulating AUG discharge \#33108, the impurities injected reach the LCFS at $t = 1.0~{\rm ms}$ after the valve trigger. Note that in this section, the time $t$ is given with respect to the time $t_{\rm inj}$ of the start of material injection. The cold gas front propagates further into the central plasma and in the process locally cools down the plasma through strong impurity radiation. As a result, the Ohmic current (with density $j_{\Upomega}$) contracts inwards where the plasma temperature has not collapsed yet (see supplementary movie). Eventually, strong current density gradients ${\rm d}j_{\Upomega}/{\rm d}\rho$ at the $q=2$ surface at $\rho = 0.7$ excite $(m,n) =(2,1)$ MHD modes and higher harmonics, thus causing rapid redistribution of heat and material inside the $q=2$ surface. In the process, the remaining plasma stored energy is dissipated globally through impurity radiation on a sub-ms time scale, decreasing the electron temperature and therefore also the plasma conductivity. Following the law of induction, strong electric fields are generated. In this environment, a seed population of runaway electrons is created due to both the hot-tail and Dreicer mechanisms. During the slower decay of the residual Ohmic current, the runaway seed population is amplified by the avalanche mechanism, establishing a significant runaway electron current at the end of the current quench, being $333~{\rm kA}$ in the simulation.

\begin{figure}
    \centering
    \includegraphics[width=104.4222mm]{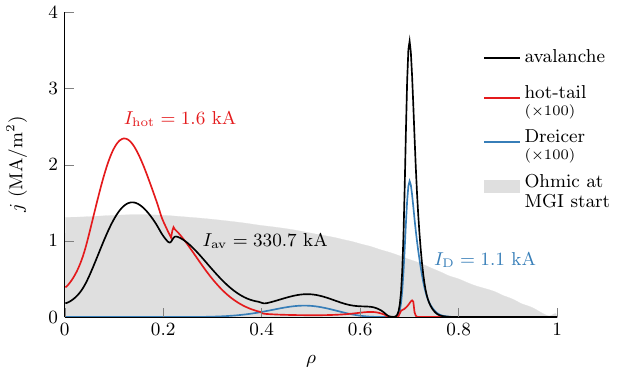}
    \caption{\label{fig_2}%
        Radial distribution of the post-disruption runaway electron current densities $j$ at the end of the current quench in simulations of AUG \#33108, generated by the avalanche mechanism (black), by the hot-tail mechanism (red), and by the Dreicer mechanism (blue). The runaway electron current densities are compared against the Ohmic current density $j_{\Upomega}$ at the start of MGI. Note, that the current densities of the hot-tail and Dreicer mechanism generated seed populations shown are multiplied by a factor of $\times 100$ given their small magnitude compared to the avalanche generated runaway current density. The spatio-temporal evolution of the runaway electron current density is additionally shown in a supplementary movie of this figure available at \href{https://arxiv.org/abs/2101.04471}{https://arxiv.org/abs/2101.04471}.
        }
\end{figure}

\subsection{The runaway seed population}
\label{sec_4.2}
Seed runaway electrons are generated due to the hot-tail and Dreicer mechanisms until the end of the thermal quench in the simulations performed. In the case of Dreicer generation, runaway occurs primarily in the vicinity of the $q=2$ surface at around $\rho = 0.7$ prior to the thermal quench (see figure~\ref{fig_2}). As the material injected begins to propagate into the plasma centre, cooling it down in the process, the Ohmic current contracts inwards to locations where the temperature has not collapsed yet. As a result, a high Ohmic current density is located in front of the cold gas, growing in magnitude as the material propagates inwards further (see supplementary movie). The maximum Ohmic current density is observed in the vicinity of the $q=2$ surface. As the cold gas front reaches this location, $(2,1)$ MHD modes are triggered. In the process, the current density is flattened inside the $q=2$ surface. As follows from the relation  
\begin{align}
    \label{eq_4.1}
	\frac{E_\parallel}{E_{\rm D}} &= \frac{m_{\rm e} \nu}{n_{\rm e} e^2} j_{\Upomega} \frac{4\upi \varepsilon_0^2 T_{\rm e}}{n_{\rm e}e^3 \ln \Lambda} = \frac{\sqrt{m_{\rm e}}}{e n_{\rm e} \sqrt{8 T_{\rm e}}} j_\Upomega ~,
\end{align}
strong electric fields $E_\parallel$ normalised to the Dreicer electric field $E_{\rm D} = n_{\rm e}e^3 \ln \Lambda/4\upi \varepsilon_0^2 T_{\rm e}$ are induced at locations of high Ohmic current density $j_{\Upomega}$, amounting to around 5\%~$E_{\rm D}$ in the vicinity of the $q=2$ surface. With the generation of Dreicer runaway electrons being, in a simple picture, exponentially sensitive to $-E_{\rm D}/E_\parallel$ \citep{Connor75}, a seed population of up to 18~kA/m$^2$ is established in the vicinity of the $q=2$ surface. 

During the thermal quench following, the electric field inside the $q=2$ surface increases up to about $3\%~E_{\rm D}$ as the rapidly decaying Ohmic current is distributed inside the $q=2$ surface more evenly. However due to the aforementioned exponential sensitivity, only a small population of additional runaway electrons is generated. Until the end of the thermal quench, the Dreicer mechanism produces a runaway current of 1.1~kA, constituting only around $0.3\%$ of the total post-disruption runaway current. Consequently, the Dreicer mechanism is relevant only for establishing a small seed population of runaways.

The hot-tail mechanism for the generation of runaways becomes important during rapid decrease of the electron temperature. Prior to the thermal quench during the inward propagation of the cold gas front, these conditions are not met. Only with the onset of the thermal quench, a noticeable population of hot-tails runaways is created inside the $q=2$ surface (see figure~\ref{fig_2}). Importantly, significant generation of more than 1~kA/m$^2$ of hot-tail current density occurs predominantly inside $\rho = 0.4$, i.e. in the region where on-axis ECRH was applied prior to MGI. At the end of the thermal collapse, a total hot-tail current of 1.6~kA is obtained, being around 0.5\% of the post-disruption runaway current. Consequently, the hot-tail mechanism also provides only a small seed population of runaways in AUG \#33108.

\begin{figure}
    \centering
    \includegraphics[width=129.8222mm]{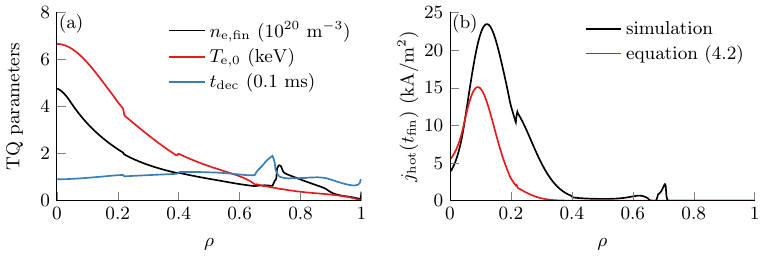}
    \caption{\label{fig_3}%
        For simulations of AUG \#33108, (a) parameters of the thermal quench, i.e. electron density at the end (black), electron temperature at the onset (red), as well as the temperature decay time scale (blue). (b) The post-quench hot-tail current density obtained through simulations (black) is compared against an analytical estimates from equation~\eqref{eq_4.2} (red).
        }
\end{figure}

The largest hot-tail current density is observed off-axis at $\rho = 0.12$, while the on-axis current density amounts to only around $\max(j_{\rm hot})/6$ despite a larger pre-quench temperature. This seemingly contradictory behaviour can be understood by evaluating the dominant contributions of the hot-tail model of equation~\eqref{eq_2.3}, thus obtaining the simplified expression (see equation~\eqref{eq_A9} of appendix~\ref{sec_A})
\begin{align}
    \label{eq_4.2}
	n_{\rm hot}^{\rm simple}(t_{\rm fin}) &= \frac{2n_{{\rm e},0}}{\sqrt{\pi}} \exp \left( -4 \left\{ \tilde{\nu} \ln\Lambda(t_0) \frac{n_{\rm e,fin} t_{\rm dec} }{T_{{\rm e},0}^{3/2}} \right\}^{2/3} \right) ~.
\end{align}
Hence in a simple estimate, the post-quench hot-tail density is exponentially sensitive to the pre-quench electron temperature $T_{\rm e,0}$, to the decay time scale $t_{\rm dec}$, and to the post-quench electron density $n_{\rm e,fin}$. Analysing the radial distribution of these quantities for AUG \#33108 (see figure~\ref{fig_3}(a)), the decay time is observed to be uniformly around $t_{\rm dec} \sim 0.1~{\rm ms}$. Therefore, the hot-tail population is predominantly determined by the ratio $n_{\rm e,fin}^{2/3}/T_{\rm e,0}$. Inside $\rho = 0.4$, this ratio decreases as the electron temperature peaks due to pre-injection ECRH. With higher pre-quench temperature, as well as due to inward impurity propagation, the impurity contributed free electron density post-quench is increased as well, peaking close to the magnetic axis, and thus increasing the ratio $n_{\rm e,fin}^{2/3}/T_{\rm e,0}$ close to the magnetic axis. As a result, the largest hot-tail population is observed off-axis.

The simple analysis following equation~\eqref{eq_4.2} is capable of reproducing the general trend of the hot-tail density obtained through evaluating the full expression~\eqref{eq_2.3} in simulations (see figure~\ref{fig_3}(b)). Consequently, the dependencies discussed are also valid for the complete model. It should be noted, that the simplified model underestimates the hot-tail density especially in the outer half radius. This behaviour occurs because factors appearing in a more general model (see equation~\eqref{eq_A7}) were simplified based on AUG disruption parameters of the central plasma. Application of the more general simplified model of equation \eqref{eq_A7} yields an estimate of the hot-tail density larger than observed in simulations.

Following this analysis, the post-quench hot-tail population observed is strongly influenced by both the pre-quench electron temperature, as well as by impurity deposition and propagation. A reduction of the hot-tail seed population can consequently be achieved by reducing the plasma temperature, slowing down the thermal quench, or depositing impurities predominantly in regions of highest temperatures. 

\subsection{Runaway electron multiplication}
\label{sec_4.3}
The vast majority of the runaway current observed in simulations of AUG \#33108, being 331~kA or 99.1\% of the post-disruption runaway current, originates from secondary runaway electrons (see figure~\ref{fig_2}), generated during knock-on collisions of thermal electrons with existing runaways from the small seed population. The radial distribution of the avalanche generated runaway current density is thus a scaled-up superposition of the seed populations. Consequently, the post-disruption runaway population is located primarily in the vicinity of the $q=2$ surface, as well as close to the magnetic axis at $\rho \sim 0.1$. Due to diffusion of the electric field during the current quench, the post-disruption runaway current density exceeds the pre-disruption Ohmic current density at these locations. As a result of impurity redistribution during the disruption and the associated impact on the evolution of the residual Ohmic current density, favourable conditions for avalanche multiplication exist predominantly close to, but inside the $q=2$ surface. Therefore, avalanche multiplication of the Dreicer generated seed population is stronger than for the hot-tail seed.

The importance of the avalanche mechanism for electron runaway in AUG \#33108 has also been observed by \citet{Insulander20} in simulations with the full-$f$ solver \texttt{CODE}. In their work, however, the small seed population was determined to consist almost entirely of hot-tail generated runaways with virtually no contribution from the Dreicer mechanism. Thus, to assess the impact of the individual source mechanisms on runaway multiplication in this framework, simulations of AUG \#33108 are repeated with only one of the primary generation mechanisms enabled.

\begin{figure}
    \centering
    \includegraphics[width=104.0694mm]{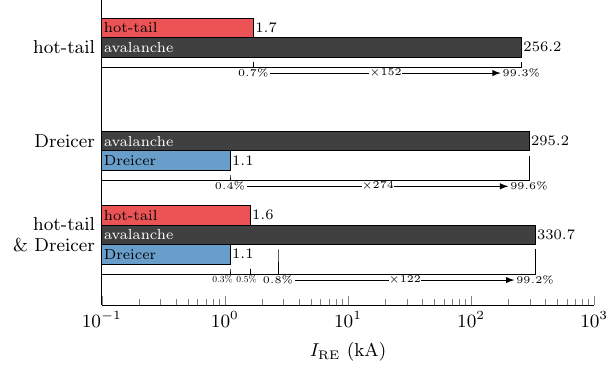}
    \caption{\label{fig_4}%
        Comparison of the post-disruption runaway current contributions from seed and avalanche mechanisms in simulations of AUG \#33108 utilizing selected source mechanisms, being 1) only the hot-tail mechanism (top), 2) only the Dreicer mechanism (middle), and 3) both the hot-tail and the Dreicer mechanism (bottom). Both the absolute runaway currents, as well as the relative strength of each generation mechanism are specified. Additionally, the avalanche multiplication factor for each simulation is listed.
        }
\end{figure}
 
In simulations considering either only the hot-tail or the Dreicer mechanism as a source for primary runaways, the seed population is reduced to 1.7~kA (-37\%) and 1.1~kA (-59\%), respectively, compared to a seed population of 2.7~kA obtained in the case of employing both mechanisms (see figure~\ref{fig_4}). Yet, the post-disruption runaway current obtained in both cases is not reduced proportionally, being 258~kA (-23\%) and 296~kA (-11\%), respectively. As the residual Ohmic current decays at similar time scales independent of the seed mechanisms employed, the post-disruption runaway current is thus determined by the avalanche multiplication time and seed population. Given the more favourable conditions for avalanche multiplication in the vicinity of the $q=2$ surface as discussed above, the smaller Dreicer generated runaway seed produces a larger secondary population than in the case utilizing only a hot-tail seed. Importantly, a comparable post-disruption runaway current is obtained in all three cases. Therefore, based on the simulations presented, the exact composition of the primary runaway seed seems to be of secondary importance in the case of AUG \#33108, as avalanche generation during the current quench dominates the dynamics.

\section{Impact of pre-disruption temperature on runaway}
\label{sec_5}
The pre-disruption electron temperature is an important parameter for hot-tail runaway electron generation during the thermal quench, as discussed in section~\ref{sec_4.2}. With increasing temperature, an exponentially increased hot-tail seed is expected to be generated. Simultaneously, increased impurity ionization is expected to occur under these conditions, potentially countering the increase of the hot-tail seed through enhanced friction. This behaviour is analysed computationally in this section by varying the pre-injection, on-axis electron temperature in the range $T_{\rm e}(\rho=0) \in [4~{\rm keV}, 20~{\rm keV}]$ in simulations of AUG \#33108.

\subsection{Set-up of electron temperature profiles}
\label{sec_5.1}
\begin{figure}
    \centering
    \includegraphics[width=133.7028mm]{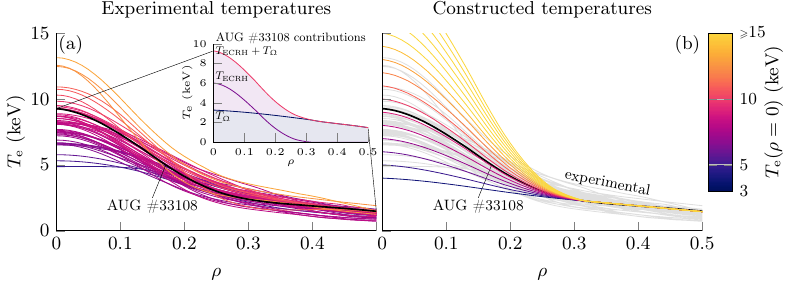}
    \caption{\label{fig_5}%
        (a) Electron temperature profiles of AUG disruption experiments similar to discharge \#33108 (see table~\ref{table_1}), constructed by Gaussian process regression using ECE and TS measurement from the last 50~ms prior to MGI. The temperature profile of AUG \#33108 can be decomposed into a contribution $T_\Upomega$ due to Ohmic heating and into a localized contribution $T_{\rm ECRH}$ due to on-axis ECRH. (b) Electron temperature profiles for the scan presented in section~\ref{sec_5} are constructed by using the profile of AUG \#33108 and scaling the ECRH contribution $T_{\rm ECRH}$, thus assuming application of varying amounts of ECRH to this baseline shot. The experimental temperature profiles of the discharges selected are shown for reference in grey. The temperature profiles are colour-coded by their on-axis values $T_{\rm e}(\rho = 0)$.
        }
\end{figure}
In AUG runaway electron experiments, on-axis ECRH is applied in the last 0.1~s prior to impurity injection to achieve high electron temperatures in the vicinity of the magnetic axis. For AUG discharges similar to AUG \#33108, the electron temperature profiles obtained through Gaussian process regression of measurements by ECE and TS thus exhibit a peaked central temperature profile of varying magnitude (see figure~\ref{fig_5}(a)). For locations around mid-radius and beyond, the local temperature and the on-axis temperature are, however, not clearly correlated. This observation motivates the approach of constructing different experimentally relevant electron temperature profiles for this investigation based on the temperature profile of AUG \#33108 under the assumption of applying varying amounts of on-axis ECRH. 

In contrast to using experimental temperature profiles of discharges with a desired pre-injection on-axis temperature, this approach ensures applying temperature profiles consistent with each other throughout the temperature range considered, thus removing the impact of peculiarities of the individual temperature profiles might have on the simulation results. Furthermore, this approach allows investigation of cases not covered (yet) experimentally, particularly at temperatures beyond 10~keV, while still ensuring experimental relevance.

To construct the temperatures profiles used, the experimental temperature profile of AUG \#33108 is separated into a contribution $T_{\Upomega}(\rho)$ due to Ohmic heating and into a contribution $T_{\rm ECRH}(\rho)$ due to ECRH (see figure~\ref{fig_5}(a)). Given the localised application, the ECRH contribution is non-vanishing only inside $\rho = 0.35$. Profiles with an arbitrary electron temperature $T_{\rm ax}$ at the magnetic axis are thus obtained by scaling the ECRH contribution, according to
\begin{align}
    \label{eq_5.1}
	T_{\rm e}(\rho) = \frac{T_{\rm ax} - T_{\Upomega}(0)}{T_{\rm ECRH}(0)} T_{\rm ECRH}(\rho) + T_{\Upomega}(\rho)~.
\end{align}
The temperature profiles constructed are consequently not modified beyond $\rho = 0.35$. The profiles used throughout this scan in the range $T_{\rm e}(\rho=0) \in [4~{\rm keV}, 20~{\rm keV}]$ are shown in figure~\ref{fig_5}(b). Compared to the temperature profiles of the discharges selected (illustrated in the same figure), the experimentally observed peaked temperature profiles are well described by the approach chosen. Therefore, the scan presented in the following describes experimentally relevant cases.

\subsection{Impact on the runaway seed}
\label{sec_5.2}
\begin{figure}
    \centering
    \includegraphics[width=136.1722mm]{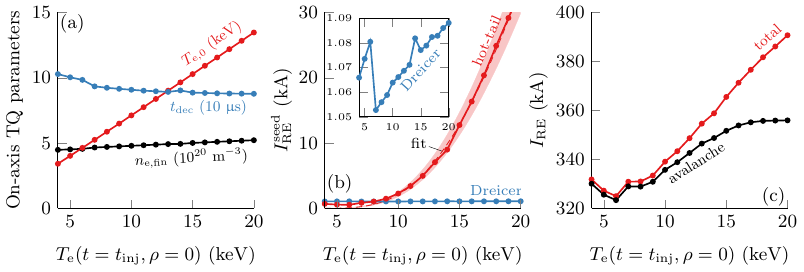}
    \caption{\label{fig_6}%
        Simulations of AUG \#33108 with increasing pre-injection on-axis electron temperature $T_{\rm e}(t=t_{\rm inj}, \rho=0)$, showing (a) on-axis thermal quench parameters, being the post-quench electron density (black), the electron temperature at the onset of the thermal quench (red), and the temperature decay time scale (blue). The runaway current obtained at the end of the disruption is shown for (b) the seed runaway population, generated by the hot-tail mechanism (red) and by the Dreicer mechanism (blue), as well as for (c) the avalanche generated runaway current (black) and the total runaway current (red). The hot-tail current in (b) is approximated by a function $I_{\rm hot}^{\rm fit} = a_0 \exp\left( - a_1/T_{\rm e,0}\right)$ (dashed red) of equation~\eqref{eq_5.2}, with fitting parameters $a_0$ and $a_1$.
        }
\end{figure}

\begin{figure}
    \centering
    \includegraphics[width=135.1138mm]{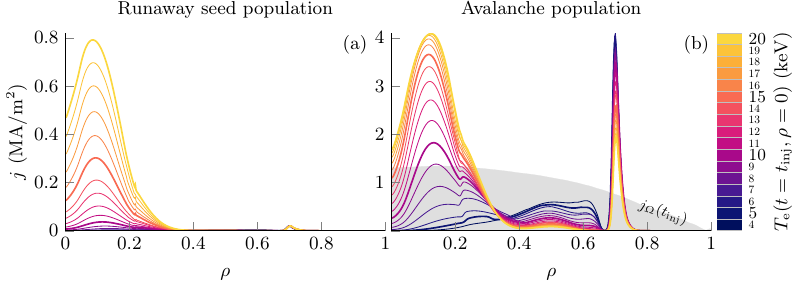}
    \caption{\label{fig_7}%
        Radial profiles of (a) the runaway electron seed current densities $j_{\rm seed}$ and (b) the post-disruption runaway electron current densities $j_{\rm av}$ generated by the avalanche mechanism in simulations of AUG \#33108 with varying pre-injection on-axis electron temperatures $T_{\rm e}(t=t_{\rm inj},\rho=0)$ ranging from 4~keV to 20~keV. For reference, the Ohmic current density $j_\Upomega$ at the start of MGI is shown (grey) in (b).
        }
\end{figure}

Increasing the pre-injection on-axis electron temperature $T_{\rm e}(t_{\rm inj},0)$ in simulations of AUG \#33108 from 4~keV up to 20~keV, the hot-tail current generated is observed to grow exponentially from a minimum value of 0.6~kA up to 33.7~kA (see figure~\ref{fig_6}(b)). For the smallest choices of $T_{\rm e}(t_{\rm inj},0)$, runaway occurs predominantly around $\rho \sim 0.7$, shifting towards $\rho \sim 0.1$ with increasing $T_{\rm e}(t_{\rm inj},0)$ (see figure~\ref{fig_7}(a)). Here, the minimum hot-tail current is observed for $T_{\rm e}(t_{\rm inj},0) = 6~{\rm keV}$. Significant hot-tail runaway eventually occurs for pre-injection on-axis temperatures beyond 10~keV (such that the hot-tail current constitutes more than 1\% of the post-disruption runaway current), generating in all cases considered a seed current density noticeably smaller than the local pre-disruption Ohmic current density $j_\Upomega$ (see figure~\ref{fig_7}). In the region of parameter space with $T_{\rm e}(t_{\rm inj},0) > 10~$keV, the hot-tail current obtained is well approximated by a function based on the simplified estimate of the hot-tail population (see equation~\eqref{eq_4.2})
\begin{align}
    \label{eq_5.2}
	I_{\rm hot}^{\rm fit}(T_{\rm e,0}(\rho=0)) &= \left( 914 \pm 58 \right) \exp \left( -4 \left\{ \frac{ \tilde{\nu} \ln\Lambda(t_0) \left\langle n_{\rm e,fin} t_{\rm dec} \right \rangle }{T_{\rm e,0}(\rho=0)^{3/2}} \right\}^{2/3} \right) ~{\rm kA} ~,
\end{align}
using the on-axis temperature $T_{\rm e,0}(\rho=0)$ at the onset of the thermal quench as dependent variable. This estimate suggests an effective spatial average of the post-quench electron density $n_{\rm e,fin}$ and decay time scale $t_{\rm dec}$ of $\left\langle n_{\rm e,fin} t_{\rm dec} \right \rangle = (1.66 \pm 0.04)\times 10^{19}~{\rm m}^{-3}~{\rm ms}$. Consequently, the hot-tail runaway current grows with increasing pre-injection temperature as anticipated following the above argument. Yet, the free electron density due to impurity ionisation does not increase at a similar rate (see figure~\ref{fig_6}(a) for the on-axis values of thermal quench parameters), due to the increasing ionisation potential of higher impurity ion charge states. At the same time, the decay time scale decreases slightly for larger temperatures, thus partially compensating the increase of the electron density. As a result, hot-tail runaway strongly increases in hotter plasmas. It should be noted that the opposite effect was observed by \citet{Aleynikov17}, where, different to the study presented in this manuscript, scenarios of instantaneous impurity deposition were investigated under the assumption of a steady-state impurity charge state distribution.

The impact of a variation of the pre-injection on-axis electron temperature on electron runaway due to the Dreicer mechanism is however negligible (see figure~\ref{fig_6}(b)), varying only by around 3\% throughout the temperature range considered. As discussed in section~\ref{sec_4.2}, Dreicer generation occurs as a result of the contracting Ohmic current predominantly in the vicinity of the $q=2$ surface, i.e. where the pre-injection temperature profile is considered unaffected by a variation of on-axis ECRH. Consequently, the Dreicer generated runaway current is approximately constant in the scenario considered.

\subsection{Runaway electron multiplication}
\label{sec_5.3}
Throughout the temperature range considered, the runaway current generated due to the avalanche mechanism increases (see figure~\ref{fig_6}(c)), but not in proportion to the strong increase of the hot-tail seed population. For pre-injection on-axis temperatures below 10~keV, the small variation of the runaway seed population results in an approximately constant avalanche current of around 330~kA. With the significant increase of the hot-tail population for larger temperatures, the avalanche generated current grows as well with temperature, yet noticeably only by a similar amount. In the range between 9~keV and 14~keV, the avalanche multiplication factor of the additional hot-tail population amounts to only between 2 and 3, even approaching a factor of 1 for $T_{\rm e,0} \to 20~{\rm keV}$. Consequently for temperatures above 17~keV, the avalanche generated current reaches a constant value of 356~kA.

The radial distribution of the avalanche current density changes throughout the range of pre-injection temperatures (see figure~\ref{fig_7}(b)) as a result of the increasing hot-tail seed population close to the magnetic axis. Occurring predominantly in the vicinity of the $q=2$ surface and around mid-radius for lower temperatures, avalanche generation shifts towards the magnetic axis to around $\rho = 0.12$. Given the large hot-tail seed in this region, significant avalanching starts earlier into the current quench, thus accelerating the decay of the residual Ohmic current.

The total runaway electron current obtained at the end of the disruption increases roughly linearly for $T_{\rm e,0} > 9~{\rm keV}$ (see figure~\ref{fig_6}(c)) due to the significantly growing hot-tail population. For the largest temperatures considered, the hot-tail seed constitutes almost 9\% of the post-disruption runaway current. Consequently, the relative impact of avalanche multiplication decreases significantly with increasing temperature, as non-negligible amounts of the finite poloidal magnetic flux available for conversion to runaways \citep{Boozer19} are consumed by a growing population of hot-tail runaways. Consequently, avalanche multiplication in future devices such as ITER may be less than predicted in previous studies \citep{Hesslow19}.

\begin{figure}
    \centering
    \includegraphics[width=136.1722mm]{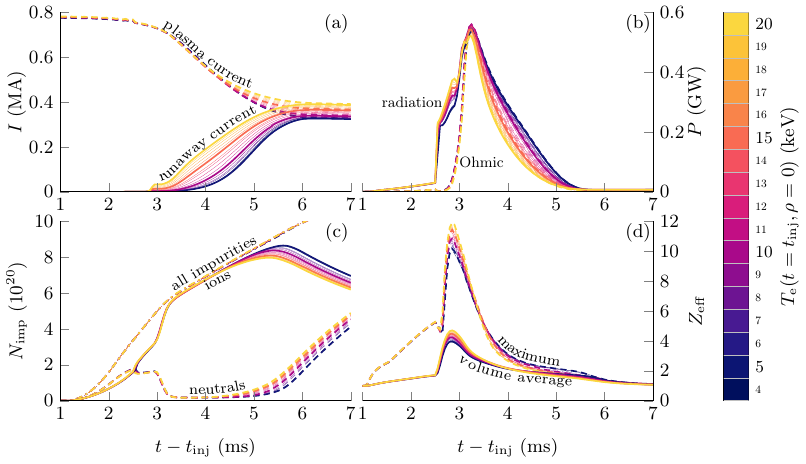}
    \caption{\label{fig_8}%
        Temporal evolution of plasma parameters throughout the disruption in simulations of varying pre-injection on-axis electron temperature $T_{\rm e}(t_{\rm inj},0)$, being (a) runaway current (solid) and plasma current (dashed), (b) Ohmic heating (dashed) and radiated power due to both line radiation and Bremsstrahlung (solid), (c) content of ionized impurities (solid), neutral impurities (dashed) and all impurities (dash-dotted), (d) volume averaged effective charge (solid) and maximum effective charge (dashed).
        }
\end{figure}

\subsection{Background plasma \& impurity evolution}
\label{sec_5.4}
The variation of the pre-injection, on-axis temperature in the simulations discussed affects not only the spatio-temporal evolution of the runaway population, but also the evolution of the background plasma and of the impurities injected (see figure~\ref{fig_8} for selected quantities). With higher initial temperature, impurity radiation (including line radiation and Bremsstrahlung) during the thermal quench is enhanced to dissipate the increased plasma thermal energy (see figure~\ref{fig_8}(b)). However, as the temperature profiles are effectively modified only in the region $\rho < 0.3$ (see figure~\ref{fig_5}(b)), making up around 10\% of the total plasma volume, the pre-injection plasma stored thermal energy increases throughout the temperature range considered only by around 30\%. The net energy lost, being the difference between Ohmic heating and impurity radiation, until the end of the thermal quench is increased by the same amount.

During the thermal quench, radiative losses far exceed 200~MW throughout the temperature range considered. As such, conductive heat transport plays a marginal role in removing heat from the central plasma. The thermal quench is therefore induced by impurity radiation in the simulations performed. The duration of the thermal quench is similar in all cases, as inferred from the occurrence of a balance between impurity radiation and Ohmic heating. This is also manifested by the temperature decay time scale (see figure~\ref{fig_6}(a)), which decreases only slightly as larger pre-injection temperatures are applied. The content of impurities inside the core plasma is identical during the thermal quench throughout the temperature range considered (see figure~\ref{fig_8}(c)). Consequently, larger densities of high impurity ionization stages are present at the end of the thermal quench in cases of high initial temperature (see figure~\ref{fig_8}(d)).

The seed population of hot-tail runaway electrons generated during the thermal quench increases in the central plasma as larger initial temperatures are applied (see section~\ref{sec_5.3}). Consequently, noticeable avalanche generation starts earlier in the disruption in the high-temperature cases (see figure~\ref{fig_8}(a)). In the process, the residual Ohmic current is depleted quicker, providing reduced amounts of Ohmic heating to the cold post-quench plasma (see figure~\ref{fig_8}(b)). As impurity radiation and Ohmic heating is balanced during the current quench, the impurities deposited in cases of hotter pre-disruption plasmas effectively recombine earlier into the current quench (see figures~\ref{fig_8}(c,d)). Nevertheless, the total length of the disruptions simulated is comparable throughout the temperature range considered, with the post-disruption runaway current being established at around 6~ms after the MGI valve trigger.

\subsection{Comparison with experimental observations}
\label{sec_5.5}
\begin{figure}
    \centering
    \includegraphics[width=103.7166mm]{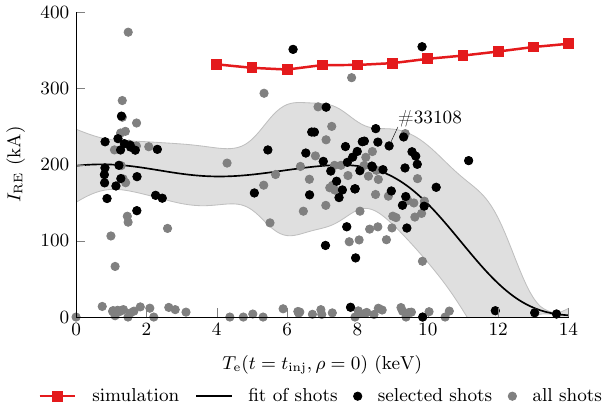}
    \caption{\label{fig_9}Post-disruption runaway electron current $I_{\rm RE}$ calculated in simulations of varying pre-injection on-axis electron temperature $T_{{\rm e},0}$ (red squares) compared to the experimental dependence $I_{\rm RE}(T_{{\rm e},0})$ of selected AUG shots similar to \#33108 (black circles) and of all runaway electron experiments performed in AUG (grey circles). Gaussian process regression of shots similar to AUG \#33108 shows the general trend observed experimentally (solid black), including uncertainties (filled grey).}
    \vspace*{-2mm}
\end{figure}

For a comparison of the post-disruption runaway current for varying pre-injection temperatures between the \texttt{ASTRA-STRAHL} simulations discussed above and AUG experiments, discharges similar to AUG experiment \#33108 are selected out of all runaway electron experiments performed in AUG. The selection is based on the pre-injection plasma current, injection quantity, toroidal magnetic field, and edge safety factor, according to the criteria listed in table~\ref{table_1}. The experimentally measured runaway current as a function of the pre-injection temperature is shown for these discharges, as well as for all runaway electron experiments performed in AUG in figure~\ref{fig_9}. The pre-injection, on-axis electron temperatures are determined applying Gaussian process regression (see section~\ref{sec_3.2}) to central ECE measurements. 

Experimental observations of the post-disruption runaway electron current as a function of the pre-injection temperature show no clear correlation between both quantities (see figure~\ref{fig_9}). Runaway currents ranging from 150~kA to 250~kA are generated regularly, but may also be as large as 350~kA or may not be observed at all. Only for temperatures well above 10~keV, electron runaway does not occur. However in this parameter region, only a small number of discharges has been performed. The experimentally observed relation between runaway current and post-injection temperature can be estimated applying Gaussian process regression. Here, a runaway current of around 190~kA independent of the temperature is on average expected for temperatures below 9~keV. 

In the \texttt{ASTRA-STRAHL} simulations of AUG discharge \#33108 performed for varying pre-injection temperatures, no strong temperature dependence of the post-disruption runaway current is observed for temperatures below 9~keV, similarly to the experimental estimate from Gaussian process regression. Yet, the calculated runaway current of around 330~kA is noticeably larger than the experimental average. Still, this behaviour is expected, as the assumption regarding the average runaway electron velocity, $\left\langle v_{\rm RE} \right\rangle = c$, may somewhat overestimate the runaway current, especially from contributions of runaways generated late into the current quench. However, a reduction of $\left\langle v_{\rm RE}\right\rangle/c$ will not proportionally reduce the post-disruption runaway current due to prolonged avalanche multiplication under these conditions (see appendix~\ref{sec_B}). As loss mechanisms for and radial transport of runaway electrons are also not considered in this work, the runaway current calculated is expected to be further overestimated. Thus, the simulations provide a pessimistic estimate of the runaway electron current. Under consideration of these effects, the relative contributions of individual generation mechanisms are expected to change only marginally, preserving the trends observed in the simulations. Importantly, neither in experiments, nor in simulations a pronounced temperature dependence of the runaway current is observed in for temperatures below 9~keV.

For temperatures above 10~keV, simulations predict a steadily increasing post-disruption runaway current, contrary to experimental observations of a vanishing runaway current. As runaway generation in all simulations occurs predominantly due to the avalanche mechanism, the absence of a post-disruption runaway current in the experiment suggests, that no seed population is present at the end of thermal quench. After all, significant avalanche multiplication of a runaway seed is expected during the current quench, given that the parallel electric field typically far exceeds the effective critical electric field under these conditions. As the amount of material injected is similar in all experiments selected, the impurity friction experienced by highly energetic electrons is assumed not to be increased. Conditions for avalanche multiplication are therefore also expected suitable for pre-injection temperatures above 10~keV.

Assuming favourable conditions for avalanche multiplication, the absence of a runaway seed population is due to either insufficient generation of primary runaways or due to the loss of the entire seed during break-up of the magnetic surfaces. However, the generation models employed in this work predict the formation of a noticeable seed population. Generation due to the Dreicer mechanism is driven by the contracting Ohmic current density in the vicinity of the $q=2$ surface. Therefore, the modification of the central electron temperature profile affects electron runaway due to momentum-space diffusion in this particular region only insignificantly, even under consideration of radial broadening of the electron temperature profile during application of increasing amounts of ECRH. In the case of hot-tail runaway, an increase of the pre-injection temperature even strongly facilitates formation of a seed population, particularly under the assumption of radial broadening of the temperature profile. Thus for pre-injection temperatures beyond 10~keV, formation of a noticeable runaway seed is also expected to occur. Note that the exponential time scale $t_{\rm dec}$ and, more importantly, the variation thereof throughout the temperature range considered cannot be determined experimentally in AUG with the available diagnostics due to insufficient temporal resolution of the TS diagnostic and the ECE signal being in high-density cutoff.

The absence of a runaway current in AUG experiments with pre-injection temperatures above 10~keV thus suggests, following the above argument, that the runaway seed is lost entirely during break-up of magnetic surfaces, i.e. before avalanche multiplication significantly increases the runaway population. Within \texttt{ASTRA-STRAHL}, this hypothesis cannot be tested as magnetic field line stochasticity and the associated runaway loss cannot be modelled self-consistently in this framework. Instead, non-linear MHD codes could be applied to investigate the existence of a transition in field line stochasticity in AUG disruptions when increasing the pre-injection temperature to above 10~keV. Similarly, the signals of magnetic diagnostics in AUG disruption experiments of varying core temperature should be analysed in future work regarding changes in MHD activity.

It should be noted, that analysing AUG runaway experiments with pre-injection temperatures above 10~keV, the low number of available discharges is not sufficient to rule out the existence of a post-disruption runaway current in this temperature region. After all, for the discharges selected, the absence of a post-disruption runaway current is also occasionally observed for temperatures below 10~keV. Therefore, further experiments with strongly increased temperatures are required to confirm or disprove the general absence of a post-disruption runaway current under these conditions.

The question of electron runaway at temperatures well above 10~keV is especially relevant for future fusion devices, such as ITER. If the runaway seed is indeed lost completely during high temperature disruptions, the risk of producing a large runaway electron current would be greatly reduced. In the opposite case, the runaway seed generated by the hot-tail mechanism is expected to contribute significantly to the overall plasma current. If, simultaneously, seed losses were to be increased (through external manipulation), poloidal flux could be removed effectively and thus avalanche multiplication hindered.

\section{Conclusion}
\label{sec_6}
In this work, runaway electron generation in ASDEX Upgrade massive gas injection experiments was investigated by means of 1.5D transport simulations performed with the coupled codes \texttt{ASTRA-STRAHL}. The suitability of this approach for the study of electron runaway in ASDEX Upgrade has recently been demonstrated by \citet{Linder20}. For this study, the toolkit chosen has been extended by a model from \citet{Smith08} describing the hot-tail population during the thermal collapse of the plasma.

Applied in simulations of argon injection in ASDEX Upgrade discharge \#33108, primary runaway generation mechanisms are calculated to both produce only a small seed population of comparable magnitude, being in total around 3~kA of fast electrons. Whereas electron runaway due to the Dreicer mechanism occurs as a result of the inward contracting Ohmic current predominantly in the vicinity of the $q=2$ surface at $\rho \sim 0.7$ prior to thermal collapse, hot-tail runaway is encountered primarily in the central plasma during the thermal quench as the post-collapse hot-tail population is exponentially sensitive to the pre-disruption temperature, $n_{\rm hot}(t_{\rm fin}) \propto \exp\left( -1/T_{\rm e,0} \right)$. At the end of the disruption, a runaway current of 331~kA is obtained in the simulations, the vast majority generated by the avalanche mechanism. A similar impact of runaway generation mechanisms has also been observed in kinetic simulations with the full-$f$ solver \texttt{CODE} \citep{Insulander20}. In the simulations presented in this work, similar post-disruption runaway electron currents are generated when neglecting one of the primary generation mechanisms. Thus, avalanche multiplication plays a significant role for the formation of a post-disruption runaway current in ASDEX Upgrade. 

Investigating the impact of varying the central electron temperature prior to argon injection in these scenarios, the post-disruption runaway current is approximately constant for on-axis temperatures below 9~keV in both simulations and experiment, generating a runaway current of around 330~kA and 190~kA, respectively. Differences are assumed to be due to the absence of runaway loss mechanisms and an overestimation of the average runaway electron velocity, $\left\langle v_{\rm RE} \right\rangle = c$. For larger temperatures up to 20~keV, simulations predict a strongly increased hot-tail population and consequently an increase of the post-disruption runaway current. Contradictorily, in the few ASDEX Upgrade discharges available in this parameter region, no post-disruption runaway current is detected. As the runaway electron models predict strong primary and secondary generation under these conditions, the absence of a post-disruption current in the experiment is considered to be caused by the loss of the entire seed population. Here, non-linear MHD codes could be applied to investigate if field line stochasticity drastically enhances seed losses. Furthermore, analysis of MHD activity inferred from measurements by magnetic diagnostics should be performed in future studies. Finally, further runaway electron experiments in ASDEX Upgrade are required to confirm or disprove the experimental trend observed.

In the simulations performed, the hot-tail mechanism provides only a small seed population of runaway electrons. Yet, the model by \citet{Smith08} employed is known to underestimate the hot-tail density \citep{Stahl16}. Under application of more elaborate (kinetic) models, hot-tail runaway is thus suspected to be significantly increased. Simultaneously, in scenarios such as ASDEX Upgrade discharge \#33108, the post-disruption runaway current is not expected to be drastically increased. However in hotter pre-disruption plasmas, a more realistic description of hot-tail runaway could provide a substantial seed population. Simultaneously, magnetic perturbations could significantly reduce avalanche multiplication \citep{Svensson21}. As this temperature range is relevant for future fusion devices, further investigation of primary runaway under these conditions is required. Here, reduced kinetic models, as e.g. being developed by \citet{Svenningsson20}, could be employed in combination with radial runaway transport coefficients, e.g. by \citet{Sarkimaki20}, to consider runaway losses.

\section*{Acknowledgements} 
The authors would like to thank M. Hoelzl for clarifying thoughts on MHD mode activity during the thermal quench. This work was supported by the EUROfusion - Theory and Advanced Simulation Coordination (E-TASC). This work has been carried out within the framework of the EUROfusion Consortium and has received funding from the Euratom research and training programme 2014-2018 and 2019-2020 under grant agreement No 633053. The views and opinions expressed herein do not necessarily reflect those of the European Commission.

\section*{Supplementary movie}
A supplementary movie is available at \href{https://arxiv.org/abs/2101.04471}{https://arxiv.org/abs/2101.04471}.

    \appendix
    
\section{Approximation of the hot-tail density}
\label{sec_A}
The temporal evolution of the hot-tail runaway electron density $n_{\rm hot}(t)$ throughout the thermal quench can be calculated with the model by \citet{Smith08}. Yet to assess the density only at the end of the thermal quench at $t_{\rm fin}$, evaluation of the model from onset of the quench at $t_0$ until $t_{\rm fin}$ is still required. Alternatively, a simple estimate can be obtained considering only the dominating contributions to the hot-tail density. 

Considering the full expression (see equation~\eqref{eq_2.3})
\begin{align}
    \label{eq_A1}
    n_{\rm hot}(t) &= \frac{4 n_{{\rm e},0}}{\sqrt{\upi} v_{{\rm th},0}^3} \int_{v_{\rm c}(t)}^{\infty} \left( v^2 - v_{\rm c}(t)^2 \right) \exp{\left( - \left[ \frac{v^3}{v_{{\rm th},0}^3} + 3 \tau(t) \right]^{2/3} \right)} {\rm d}v ~,
\end{align}
the velocity distribution function is evaluated beyond the critical velocity for runaway $v_{\rm c}^2 = e^3 n_{\rm e} \ln\Lambda/4\pi\varepsilon_0^2 m_{\rm e} E_\parallel$.  Throughout the process of thermal collapse, an initially large $v_{\rm c}(t_0) \gg v_{\rm th,0}$ will eventually approach $v_{\rm th,0}$, i.e. $v_{\rm c}(t_{\rm fin}) \to v_{\rm th,0}$, as the local electric field $E_\parallel$ strongly increases. Simultaneously, the parameter $\tau(t)$ grows throughout the quench according to $\tau(t) = (t - t_{\rm dec}) \nu_0 n_{\rm e,fin}/n_{\rm e,0}$. Consequently, the exponent of the exponential function of equation~\eqref{eq_A1} starts typically far from unity, i.e. $v_{\rm c}^3/v_{\rm th,0}^3 + 3\tau \gg 1$. As the exponential function decreases rapidly for $v > v_{\rm c}$, electrons with $v \gtrsim v_{\rm c}$ contribute dominantly to the velocity space integral. Under these assumptions, the argument of the exponential function in equation~\eqref{eq_A1} can be approximated as
\begin{align}
    \label{eq_A2}
	- \left[ \frac{v^3}{v_{{\rm th},0}^3} + 3 \tau(t) \right]^{2/3} &\approx - \left( \frac{v}{v_{\rm c}(t)} \right)^2 \left[ \left( \frac{v_{\rm c}(t)}{v_{\rm th,0}} \right)^3 + 3 \tau(t) \right]^{2/3} ~.
\end{align}
The solution of the velocity space integral is thus readily obtained as
\begin{align}
    \label{eq_A3}
	n_{\rm hot}(t) &= \frac{2n_{\rm e,0}}{\sqrt{\pi}} \frac{\left( \frac{v_{\rm c}(t)}{v_{\rm th,0}} \right)^3}{\left[ \left( \frac{v_{\rm c}(t)}{v_{\rm th,0}} \right)^3 + 3 \tau(t) \right]^{2/3}} \exp\left(- \left[ \left( \frac{v_{\rm c}(t)}{v_{\rm th,0}} \right)^3 + 3 \tau(t) \right]^{2/3} \right) ~.
\end{align}
A similar approximation is derived in \citet{Smith08}, where the numerator of the pre-exponential fraction is amended by $+ 3\tau(t)$. However, the resulting expression overestimates the hot-tail population, compared to evaluation of equation~\eqref{eq_A1}.

To obtain the post-quench hot-tail population, the above expression has to be evaluated at time $t^\star$, when the exponent of the exponential function reaches its maximum value. This in turn requires evaluation of the temporal evolution of the critical velocity $v_{\rm c}(t)$. Considering an exponential decay of the electron temperature in the limit $T_{\rm e,fin} \ll T_{{\rm e},0}$, i.e. $T_{\rm e}(t) = T_{\rm e,fin} + (T_{\rm e,0} - T_{\rm e,fin}) e^{-t/t_{\rm dec}} \approx T_{\rm e,0} e^{-t/t_{\rm dec}}$, the evolution of the electric field required for $v_{\rm c}(t)$ is obtained as
\begin{align}
    \label{eq_A4}
	E_\parallel(t) = \frac{j_{\Upomega,0}}{\sigma(t)} = \frac{e^2 \sqrt{m_{\rm e}}\ln\Lambda}{8\sqrt{2}\pi \varepsilon_0^2 T_{\rm e}(t)^{3/2}} j_{\Upomega,0} = E_{\parallel,0} \exp \left( \frac{3}{2} \frac{t}{t_{\rm dec}} \right) ~,
\end{align}
with $\sigma$ being the plasma conductivity. Note, that the Ohmic current density $j_{\Upomega}(t)$ is in good approximation constant throughout the part of the thermal collapse relevant for hot-tail runaway. Alternatively, the evolution of the electric field can be evaluated through $\frac{{\rm d}}{{\rm d}t} \left\{ \sigma(t) E_\parallel(t) \right\} = - 2 R E_\parallel(t)/La^2$ \citep{Hesslow18}, with major radius $R$, minor radius $a$, and inductance $L$. Using typical AUG parameters, deviations with respect to expression \eqref{eq_A4} become important only for $t/t_{\rm dec} \gtrsim 5$, i.e. when the hot-tail population is already established. Thus, the ratio of velocities can be written as
\begin{align}
    \label{eq_A5}
	\frac{v_{\rm c}(t)}{v_{\rm th,0}} &= \sqrt{\frac{e n_{\rm e,0}}{j_{\Upomega,0}} \sqrt{\frac{2 T_{\rm e,0}}{m_{\rm e}}}} \exp \left( - \frac{3}{4} \frac{t}{t_{\rm dec}} \right) = \frac{v_{\rm c,0}}{v_{\rm th,0}} \exp\left(-\frac{3}{4} \frac{t}{t_{\rm dec}} \right)
\end{align}
and thus the exponent of the exponential function obtains its maximum value at time (see also \citet{Smith08})
\begin{align}
    \label{eq_A6}
	t^\star = \frac{4}{9} t_{\rm dec} \left\{ 3 \log \left( \frac{v_{\rm c,0}}{v_{\rm th,0}} \right) - \log\left( \frac{4}{3} \nu_0 \frac{n_{\rm e,fin}}{n_{\rm e,0}} t_{\rm dec} \right) \right\} ~.
\end{align}
\begin{figure}
    \centering
    \includegraphics[width=108.6556mm]{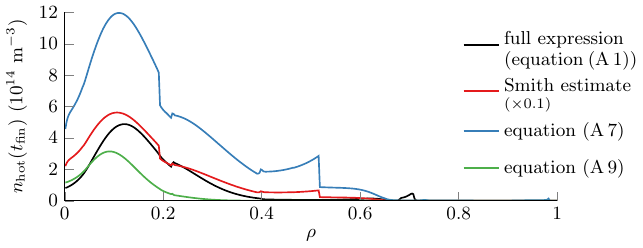}
    \vspace*{-1mm}
    \caption{\label{fig_10}%
        Hot-tail population of AUG \#33108 calculated in simulations of \texttt{ASTRA-STRAHL} evaluating the full expression of equation~\eqref{eq_A1} (black), see section~\ref{sec_4}, compared to analytical estimates from equation (23) of \citet{Smith08} (red), equation~\eqref{eq_A7} (blue), and equation~\eqref{eq_A9} (green). Note, that the estimate using the approximation by \citet{Smith08} is scaled by a factor of $\times 0.1$.
        }
    \vspace*{-2mm}
\end{figure}
Applied in equation~\eqref{eq_A3}, the hot-tail population at the end of the thermal quench can be estimated as
\begin{align}
    \label{eq_A7}
	n_{\rm hot}(t_{\rm fin}) \approx & \frac{2 n_{\rm e,0}}{\sqrt{\pi}} \frac{ \mathcal{F}^{1/3}}{\mathcal{G}^{2/3}} \exp \left( - \left( \mathcal{F} \mathcal{G} \right)^{2/3} \right) ~,
\end{align}
where
\begin{align}
    \label{eq_A8}
	\mathcal{F} &= \frac{4}{3} \nu_0 \frac{n_{\rm e,fin}}{n_{\rm e,0}} t_{\rm dec} ~, &
	\mathcal{G} &= 3 \log\left( \frac{v_{\rm c,0}}{v_{\rm th,0}} \right) - \log \mathcal{F} - \frac{5}{4} ~.
\end{align}

The radial variation of expression~\eqref{eq_A7} is primarily determined by the contribution of $\mathcal{F}$ in the exponent of the exponential function. To emphasize this dependence, further simplifications can be made by applying values for the disruption parameters as typically occurring in AUG disruption experiments, yielding a pre-exponential factor of $\mathcal{F}/\mathcal{G}^2 \approx 1$ and inside the exponent $(4 \mathcal{G}/3)^{2/3} \approx 4$. Writing the collision frequency as $\nu_0 n_{\rm e,fin}/n_{\rm e,0} \equiv \tilde{\nu} \ln\Lambda(t_0) n_{\rm e,fin}/T_{\rm e,0}^{3/2}$, the hot-tail runaway density can thus be expressed as
\begin{align}
    \label{eq_A9}
	n_{\rm hot}^{\rm simple}(t_{\rm fin}) &= \frac{2 n_{\rm e,0}}{\sqrt{\pi}} \exp\left(  - 4 \left\{ \tilde{\nu} \ln\Lambda(t_0) \frac{ n_{\rm e,fin} t_{\rm dec}}{T_{\rm e,0}^{3/2}} \right\}^{2/3} \right) ~.
\end{align}
Albeit being a simple estimate for the hot-tail population, key dependencies on thermal quench parameters are readily clear evaluating this expression. Estimates of the hot-tail population for the simulation of AUG discharge \#33108 presented in section~\ref{sec_4} are shown in figure~\ref{fig_10} for the different analytical expressions introduced. The simple estimate of equation~\eqref{eq_A9} agrees rather well with the hot-tail population obtained through evaluation of the full expression~\eqref{eq_A1}, thus illustrating the suitability of equation~\eqref{eq_A9} to assess the dependence of the hot-tail population on parameters of the thermal quench.

\section{Average runaway electron velocity}
\label{sec_B}
In the simulations presented, the runaway electron current density $j_{\rm RE}$ is calculated from the number density $n_{\rm RE}$ under the assumption, that runaway electrons travel with the speed of light, i.e. the average runaway electron velocity $\left\langle v_{\rm RE} \right\rangle = c$. For large kinetic energies $E_{\rm kin} > 6.1 m_{\rm e}c^2 = 3.1~{\rm MeV}$, this gives a less than 1\% error. However, the validity of this assumption is often questioned. Therefore, it is demonstrated in this section, that a reasonable choice of $\left\langle v_{\rm RE} \right\rangle \sim c$ has only a minor impact on the amount of post-disruption runaway current generated. 

\begin{figure}
    \centering
    \includegraphics[width=98.4250mm]{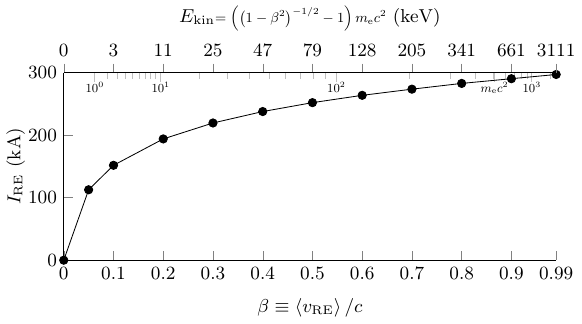}
    \caption{\label{fig_11}%
        Post-disruption runaway electron current $I_{\rm RE}$ in simulations of AUG \#33108 applying a varying average runaway electron velocity $\left\langle v_{\rm RE}\right\rangle$. For reference, the corresponding kinetic electron energy $E_{\rm kin}$ is given. Note, that these simulations were carried out with decreased temporal resolution for illustrative purposes.
        }
\end{figure}

In simulations of AUG \#33108 of varying average runaway electron velocity, the post-disruption runaway electron current obtained is rather insensitive to a moderate modification of $\left\langle v_{\rm RE}\right\rangle$, as shown in figure~\ref{fig_11}. Note, that these simulations presented were performed with decreased temporal resolution for illustrative purposes. Assuming an average velocity of 50\%~c, the post-disruption runaway current is reduced by 15\%. Only for $\left\langle v_{\rm RE} \right\rangle \lesssim 20\%~c$, the runaway current calculated falls off quickly. Importantly, the associated kinetic energy of the runaway electrons is well below $m_{\rm e}c^2$ for both choices of $\left\langle v_{\rm RE} \right\rangle/c$ discussed, approaching even the pre-disruption thermal electron energy. However as the bulk of the runaway electron population is expected to reach kinetic energies above the rest mass energy, corresponding to $\left\langle v_{\rm RE} \right\rangle > 87\%~c$, the post-disruption runaway current is not significantly affected by a variation of $\left\langle v_{\rm RE} \right\rangle$ within these bounds. 

A variation of the average runaway electron velocity can also be considered as a variation of the strength of primary runaway electron generation under the assumption $\left\langle v_{\rm RE} \right\rangle = c$. Writing $\beta \equiv \left\langle v_{\rm RE}\right\rangle/c$, the macroscopic transport equation~\eqref{eq_2.1} for the primary runaway electron current density can be expressed as (neglecting radial transport)
\begin{align}
    \label{eq_B1}
	\frac{\partial j_{\rm seed}}{\partial t} &= e \left\langle v_{\rm RE} \right\rangle S_{\rm seed} = e \left( c \beta \right) S_{\rm seed} = e c \left(\beta S_{\rm seed}\right) ~,
\end{align}
hence describing either a variation of $\left\langle v_{\rm RE}\right\rangle$ or of $S_{\rm seed}$. Simultaneously, avalanche multiplication is described by
\begin{align}
    \label{eq_B2}
	\frac{\partial j_{\rm av}}{\partial t} &= e \left\langle v_{\rm RE} \right\rangle n_{\rm RE} \tilde{S}_{\rm av} = \left( j_{\rm av} + j_{\rm seed} \right) \tilde{S}_{\rm av} ~,
\end{align}
thus not explicitly considering assumptions regarding the average runaway electron velocity. A variation of $\left\langle v_{\rm RE}\right\rangle$ does therefore directly affect only the primary population. Hence, varying $\beta$ within the range $\left[ 0, 1 \right]$, the source strength $S_{\rm seed}$ can be considered scaled by this factor instead of $\left\langle v_{\rm RE}\right\rangle$. Under these conditions of decreased primary generation, however, avalanche multiplication is not affected proportionally, as discussed in section~\ref{sec_4.3}.

    \bibliographystyle{jpp_with_hyperlinks}
    \bibliography{ms}

\end{document}